# Topological Valley Currents in Bilayer Graphene/Hexagonal Boron Nitride Superlattices

Kosuke Endo[1,2*], Katsuyoshi Komatsu[1*], Takuya Iwasaki[3], Eiichiro Watanabe[4], Daiju Tsuya[4], Kenji Watanabe[5], Takashi Taniguchi[5], Yutaka Noguchi[2], Yutaka Wakayama[1], Yoshifumi Morita[6†], and Satoshi Moriyama[1†]

[1]International Center for Materials Nanoarchitectonics (WPI-MANA), National Institute for Materials Science (NIMS), Tsukuba, Ibaraki 305-0044, Japan.
[2]School of Science & Technology, Meiji University, Kawasaki 214-8571, Japan.
[3]International Center for Young Scientists (ICYS), NIMS, Tsukuba, Ibaraki 305-0044, Japan.
[4]Nanofabrication Platform, NIMS, Tsukuba, Ibaraki 305-0047, Japan.
[5]Research Center for Functional Materials, NIMS, Tsukuba, Ibaraki 305-0044, Japan.
[6]Faculty of Engineering, Gunma University, Kiryu, Gunma 376-8515, Japan.

* These authors contributed equally to this work.

† Corresponding author. Email: morita@gunma-u.ac.jp; MORIYAMA.Satoshi@nims.go.jp

**Abstract**
Graphene superlattices have recently been attracting growing interest as an emergent class of quantum metamaterials. In this paper, we report the observation of nonlocal transport in bilayer graphene (BLG) superlattices encapsulated between two hexagonal boron nitride (hBN) layers, which formed hBN/BLG/hBN moiré superlattices. We then employed these superlattices to detect a long-range charge-neutral valley current using an all-electrical method. The moiré superlattice with broken inversion symmetry leads to a "hot spot" at the charge-neutral point (CNP), and it harbors satellites of the CNP. We observed nonlocal resistance on the order of 1 kΩ, which obeys a scaling relation. This nonlocal resistance evolves from an analog of the quantum Hall effect but without magnetic field/time-reversal symmetry breaking, which is associated with a hot-spot-induced topological valley current. This study should pave the way to developing a Berry-phase-sensitive probe to detect hot spots in gapped Dirac materials with inversion-symmetry breaking.

_______________________________

Graphene superlattices are an emergent class of quantum metamaterials with considerable promise. In this letter, we explore the transport properties of bilayer graphene (BLG) superlattices, focusing on the topological current associated with a valley degree of freedom. A valley in an energy band implies a degenerate local minimum (maximum) in the conduction (valence) band and is referred to as *K* and *K'* in the case of graphene, which correspond to valley pseudospin states [1]. Graphene is associated with two valley pseudospin states, which transform into each other under spatial inversion. The broken inversion symmetry induces valley-contrasted physics through the emergence of a "hot spot" for each valley, which leads to the generation of valley Hall currents owing to the finite Berry curvature in the energy band [1,2]. The nonlocal transport is a Berry-phase-sensitive probe to detect such a hot spot in graphene superlattices [3, 4, 5]. More recently, quantum valley current was detected through the nonlocal resistance in the quantum limit [5]. This is in analogy with quantum Hall effect but without magnetic-field/time-reversal symmetry breaking, which is associated with hot-spot-induced

"topological current" [6, 7, 8]. The BLG analog of the valley current has also been observed without a superlattice structure [9, 10], wherein Berry curvature is induced through an electric field due to an external gate. The Berry curvature plays the role of a pseudomagnetic field in momentum space, and its sign reverses between the two valleys [1, 2]. In addition, it leads to the generation of a transverse valley current. Combined with the inverse valley Hall effect, which converts the valley current into transverse electric fields, we can detect the valley current using an all-electrical method. In the above picture, the nonlocal transport was meditated by conduction because of bulk valley Hall current. Moreover, possible scenarios were proposed due to edge-channel conduction in reference [5]. For a theoretical approach on the edge picture, see also reference [11].

In this study, we employed hexagonal boron nitride (hBN)/BLG/hBN moiré superlattices to detect the topological valley current due to the hot spot using an all-electrical method in a four-terminal H-bar device. In the moiré superlattices, a long-wavelength moiré pattern develops, which leads to Hofstadter's butterfly under a magnetic field [12, 13], wherein the hot spot emerges because of broken inversion symmetry. The electrical current injected at one end of the H-bar induces a topological valley current in the transverse direction because of the hot spot, and this current is converted to a voltage drop at the other end of the H-bar through the inverse effect [3, 4] (see also Figures 2(a)).

Recently, graphene superlattices due to the moiré structure have been intensively studied. In particular, since the work of Bistritzer and MacDonald [14, 15, 16], moiré bands/butterflies in BLG have been extensively studied, which are analogous to those in single-layer graphene (SLG) [12]. BLG has a unique electronic structure quite different from that of SLG, and its band gap can be tuned by applying an electric field perpendicular to the system. The two layers in BLG generally exhibit AB (Bernal) stacking. Such AB-stacked BLG on an hBN system has a remarkable character. The hBN leads to an energy gap and hot spot due to broken inversion symmetry. Although the detailed stacking/electronic structure needs more careful analysis [11,16], the presence of a hot spot should be protected topologically, and it guarantees the emergence of topological current. Specifically, broken inversion symmetry is induced by placing the BLG on an hBN substrate with a precise alignment angle $\theta$ near ~0°. A long-wavelength (~14 nm) moiré pattern develops, which stems from the small lattice mismatch between BLG and hBN [13]. The moiré superlattice leads to an energy gap/hot spot at the charge-neutral point (CNP), and satellites of the CNP emerge. The moiré superlattice also induces an intriguing energy spectrum under a magnetic field, which is referred to as "Hofstadter's butterfly" [13]. When subject to a magnetic field, the resistance peaks lead to first- and second-generation Landau fans. The first generation corresponds to the CNP, and the second generation is due to inversion-symmetry breaking and corresponds to the satellites of the CNP. To detect the topological valley current, we observe the nonlocal resistance in our BLG superlattices. Observing such resistance is thus a way to probe the hot spot and associated topological current. We also confirm the cubic scaling relation between the nonlocal resistance and the longitudinal resistance, which indicates the emergence of bulk topological current.

To fabricate the superlattice heterostructure, we conducted the dry transfer method described in Ref [17]. First, using the mechanical exfoliation method on Kish graphite, graphene flakes were transferred onto a $SiO_2$/Si substrate covered with poly(methyl methacrylate) (PMMA) on top and a water-soluble layer (polyacrylic acid: PAA) in between. BLG was identified by the optical contrast between the flake and substrate. By utilizing the method same as that applied for the graphene, hBN flakes were prepared using a different substrate. PAA layer was dissolved in water for the hBN substrate; the floating PMMA layer with the hBN flake was transferred onto a polymer stamp made of polydimethylsiloxane. The substrate with the BLG flake and stamp with hBN flake were placed on a home-made transfer system facing each other but slightly separated. The temperature of the substrate stage was set to 60°C. The sample stage was rotated by carefully aligning the edge of each flake, i.e., the edge at which a crystal face appears [18]. The flakes were brought closer to each other until they were briefly touching. As a result, the hBN flake picked up the BLG flake, and hence, both flakes were on the stamp. Next, the hBN/BLG was dropped onto an hBN flake on the $SiO_2$ (300 nm)/Si substrate using the same method, thus obtaining hBN/BLG/hBN. Pick-up and drop-down were controlled by the relationship between the flake size, i.e., a larger flake picks up a smaller flake. By utilizing atomic force microscopy, we obtained the measurement of the thicknesses of both the top and bottom hBN layers, i.e., 30 nm, where the aligned bottom hBN is set to induce a superlattice structure. To pattern the hBN/BLG/hBN into the H-bar geometry, we conducted an electron beam (EB) lithography, which was followed by reactive ion etching using $CHF_3/O_2/N_2$ plasma. To fabricate the one-dimensional contact electrodes [19], we conducted EB lithography, EB deposition with Cr/Au (5 nm/55 nm), and lift-off.

The longitudinal and nonlocal resistances are defined by applying the four-terminal resistance, $R_{ij,kl}$, which can be defined by the voltage between the terminals $k$ and $l$ divided by the electrical current injected between the terminals $i$ and $j$ (see the inset of Figures 1(a) for the definition of the terminals). These resistances were measured by utilizing a low frequency (17 Hz) lock-in technique with an AC current excitation of 1–10 nA. $R_{12,43}$ is the longitudinal resistance, and $R_{14,23}$ is the nonlocal resistance. It is to be noted that our device illustrates four-terminal rectangular structures in a quasiballistic regime, wherein the mean-free path is close but within the dimensions of the device, which is of the order 1 μm, and the current–voltage (I–V) characteristics demonstrated Ohmic behavior [20]. As a function of the magnetic field $B$, $R_{13,42}$ leads to the Hall coefficient after a symmetrization [21, 22].

Figures 1(a) and 1(b) show our H-bar devices, in which BLG is encapsulated between two hBN layers. Figure 1(c) shows the longitudinal resistance ($R_{xx} = R_{12,43}$) mapping plot as a function of the back-gate voltage ($V_g$) and perpendicular magnetic field ($B$) at 10 K, exhibiting the spectrum called Hofstadter's butterfly. The $R_{xx}$ and Hall resistances ($R_{xy} = R_{13,42}$) exhibit basically the same pattern as those in previous reports [13]; specifically, a pronounced peak appears in the longitudinal resist at the CNP at $V_g \sim -3.0$ V, and satellite resistance peaks appear at $V_g \sim +30.0$ V and $-34.4$ V, which we call satellites for simplicity. From the position of the satellite peaks, we estimate the alignment angle $\theta$ between the BLG and hBN of less than ~0.2°, as in Ref [12,13]. When $B$ is applied, these two resistance peaks transform into the first- and second-generation Landau fans, respectively, as shown in Fig. 1(c). Near the central CNP, the Hall resistance changes its sign as the $V_g$ is swept. The same trend is observed near the two satellite peaks. The plot of $\sigma_{xy}$ (= $R_{xy}/(\rho_{xx}^2 + R_{xy}^2)$, where $\rho_{xx} = R_{xx}(W/L)$) as a function of $V_g$ shows a Landau level formation with plateau steps of $4e^2/h$, which reflects the nature of BLG, as shown in Fig. 1(d).

Next, we discuss the nonlocal measurement results. The electrical current injected at one end of the H-bar induces a topological valley current owing to the presence of hot spot in the transverse direction, which is converted to a voltage drop at the other end through the inverse effect (see also Figure 2(a)). Figure 2(b) shows the observation of nonlocal transport in the hBN/BLG/hBN superlattices as a function of the carrier density ($n$) near the CNP. The carrier density induced by the gate voltage is estimated from the Hall effect measurement at low magnetic fields, which is $6.5 \times 10^{10}$ cm$^{-2}$/V. This value is agreement with the model estimation for the surface charge density induced by the field effect in the device structure of Fig. 1(b). We observed nonlocal resistance ($R_{nl} = R_{14, 23}$) on the order of 1 kΩ at the CNP. Note that our device is of high quality, as the upper bound of the charge inhomogeneity at the CNP is estimated to be $\delta n$ ~$2.8 \times 10^{10}$ cm$^{-2}$ from the peak width of the $\rho_{xx}$ of CNP, as in Ref [23]. According to the analysis described in [21,22], the Ohmic contribution is an order of magnitude smaller than $R_{nl}$. Therefore, the Ohmic contribution can be ruled out as a main factor influencing our observation. Herein, some comments are appropriate for the satellites. This paper focuses on nonlocal transport at the CNP. In the devices considered herein, a nonlocal response at the satellites is suppressed [3]. Even though the mechanism for the intensity of nonlocal resistance is yet to be clarified theoretically (see, for example, reference [11]), it is empirically known that the quality of devices is connected to it. Moreover, the nonlocal response is enhanced because of the ballistic character of the devices [5]. Although the devices applied herein are of high quality, they are still considered in a quasiballistic regime. As demonstrated in Figure 2(c), the cubic scaling relation between $\rho_{xx}$ and $R_{nl}$ signifies the emergence of a bulk topological current at a temperature of 6 K (and higher temperature regime) with a minor deviation because of the Ohmic contribution [3, 24, 25]. However, a large deviation from such scaling was detected at a temperature of 1.5 K, as depicted in the inset of Figure 2(c). This is the same as in the following reference [5], where it was assigned to the quantum limit/possible emergence of edge-channel conduction [11]. Owing to the quasiballistic character of the devices applied in this paper, we observe that our devices are still not in the genuine quantum limit but a precursor was detected. Figure 2(d) shows the temperature dependence of $\rho_{xx}$ and $R_{nl}$. The data in the high-temperature region are fitted by the thermally activated transport model i.e. Arrhenius law $\rho_{xx}$ (and $R_{nl}$) in the form of $\exp(A/2k_BT)$, where $A$, $k_B$, and $T$ are the activation energy, Boltzmann constant, and temperature, respectively. As the $A$ value of ~51 K (~4.4 meV) for the nonlocal transport is approximately three times higher than that for the local transport of ~20 K (~1.7 meV), this result further supports the consistency of our picture [26]. Note that this contradicts the claim in reference [3], which is inconsistent with the bulk-conduction picture. It deviates from the simple Arrhenius law if the temperature is further lowered. There are several scenarios which are yet to be settled. A scenario is disordered effects. Even though such effects can lead to a variable-range hopping model, it is clear that it cannot be applied to all the low-temperature regions. The saturation of the low-temperature regime has also been observed, for e.g., in ref. [5], where it is assigned to be the quantum limit (see also ref. [11] for a recent theoretical approach).

Figures 3(a) and 3(b) illustrates the local and nonlocal resistance mapping plots as a function of $V_g$ and $B$ at 30 K, in which the nonlocal resistance is suppressed at $B = 0$ T. The fade-out of the nonlocal resistance at 30 K can also be observed in the Arrhenius plot, as depicted in Figure 2(d). Therefore, we note that the Hofstadter's butterfly spectrum for both the local and nonlocal resistances appear under a finite magnetic field $B$, which has a different origin from the valley current [5]. Under a high magnetic field, i.e., quantum Hall regime, edge-channel conduction should play a dominant role. Furthermore, a bulk spin and a heat current can also play some roles. A careful assignment is left as a future work.

In summary, we observed nonlocal transport in BLG superlattices. Nonlocal resistance appeared at the CNP on the order of 1 kΩ, satisfying the scaling law, which indicates the emergence of long-range bulk topological valley currents in the BLG superlattices. "Higher-generation" Dirac fermions have been proposed to emerge in several solids, including the recently fabricated both twisted [27] and non-twisted [28] BLG superlattices. If the inversion and/or time-reversal symmetry is broken in those superlattices due to, for example, substrate/many-body effects, this symmetry breaking should lead to the mass generation of Dirac fermions and should harbor hot spots, as observed in this paper. Nonlocal transport in such systems is an area of future study. Notably, this valley Hall state should play the role of a "parent state" when doped, such as through electrostatic tuning.

**Acknowledgments:**

The device fabrication and measurement were supported by the Japan Society for Promotion of Science (JSPS) KAKENHI 26630139, and the NIMS Nanofabrication Platform Project, the World Premier International Research Center Initiative on Materials Nanoarchitectonics, sponsored by the Ministry of Education, Culture, Sports, Science and Technology (MEXT), Japan. S.M. acknowledges financial support from a Murata Science Foundation. Growth of hBN crystal was supported by the Elemental Strategy Initiative conducted by the MEXT, Japan, and the CREST (JPMJCR15F3), JST.

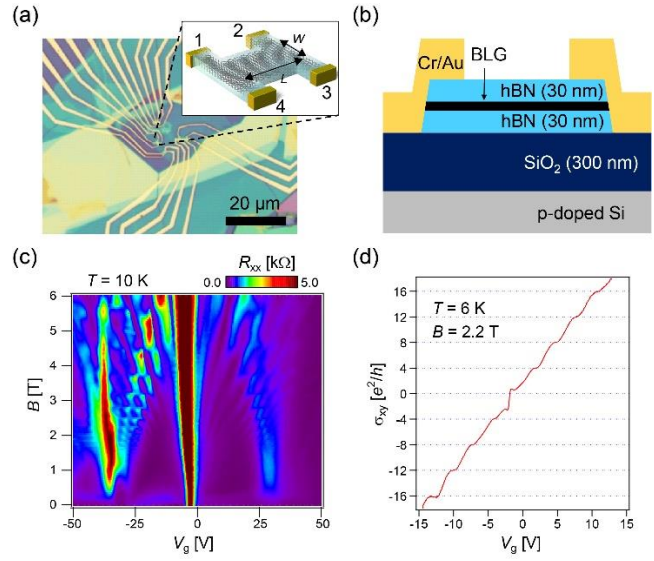

**Figure 1: Device structure and characterization of the hBN/BLG/hBN superlattices.**
(a) Optical image of our BLG superlattices. The inset schematically shows the H-bar device structure with the characteristic length of $L = 1.5$ µm and $W = 1.0$ µm.
(b) Schematic cross-section of the BLG superlattice device.
(c) Intensity mapping plot of the longitudinal resistance $R_{xx}$ as a function of $V_g$ and a perpendicular magnetic field $B$ at $T = 10$ K.
(d) Quantum Hall effect (QHE) at $T = 6$ K and $B = 2.2$ T. Plateaus with quantized values signify the QHE characteristics of the BLG.

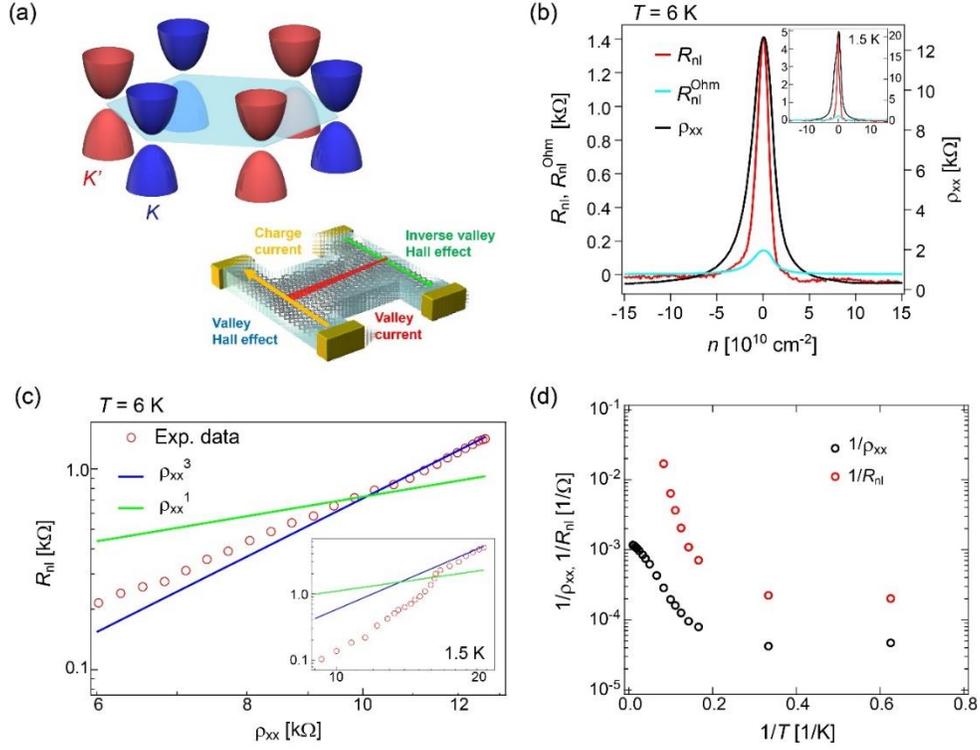

**Figure 2: Nonlocal resistance and its scaling in the hBN/BLG/hBN superlattices.**
(**a**) Schematic picture of the energy band for the hBN/BLG/hBN superlattices and valley currents.
(**b**) Nonlocal resistance $R_{nl}$ with the Ohmic contribution $R_{nl}^{Ohm}$ and local resistivity $\rho_{xx}$ as a function of the carrier density at $T = 6$ K near the CNP. Inset: the same at $T = 1.5$ K.
(**c**) Scaling relation between the local resistivity $\rho_{xx}$ and the nonlocal resistance $R_{nl}$ at 6 K. Circle symbols show the experimental data. Blue and green lines correspond to the scaling for $\rho_{xx}^3$ and $\rho_{xx}^1$, respectively. The cubic scaling relation between $\rho_{xx}$ and $R_{nl}$ indicates the emergence of a bulk topological current with a minor deviation due to the Ohmic contribution. Inset: the same at $T = 1.5$ K.
(**d**) Arrhenius plot of the inverse of $\rho_{xx}$ and $R_{nl}$ at the CNP.

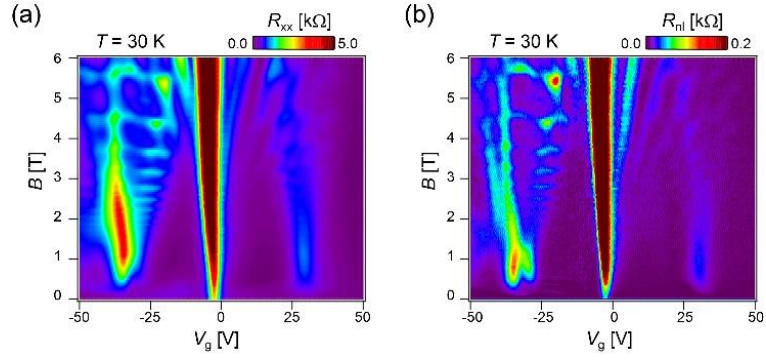

**Figure 3: Fade-out of the nonlocal transport behavior in the hBN/BLG/hBN superlattices at 30 K.**
(a) Intensity mapping plot of the local resistivity $R_{xx}$ as a function of $V_g$ and a perpendicular magnetic field $B$ at 30 K.
(b) Intensity mapping plot of the nonlocal resistance $R_{nl}$ as a function $V_g$ and $B$ at 30 K.